\documentclass[11pt,a4paper]{article}
\usepackage{amsmath}
\usepackage{amssymb}
\usepackage{amsthm}
\usepackage{epsfig}
\usepackage{times}
\usepackage{wrapfig}
\usepackage{rotating}

\usepackage[hang,stable]{footmisc}

\newcommand{\reals}{\mathbb{R}}
 
\newcommand{\matr}{\mathbf}
\newcommand{\lieg}{\mathrm}
\newcommand{\q}{\mathbf{q}}
\newcommand{\h}{\mathbf{g}}

\begin{document}
\ifx\href\undefined\else\hypersetup{linktocpage=true}\fi

\title{Instants in physics\\
-- point mechanics and general relativity --}

\author{Domenico Giulini            \\
        Institute for Theoretical Physics\\
        Riemann Center for Geometry and Physics\\
        Leibniz University Hannover\\ 
        Appelstrasse 2, D-30167 Hannover, Germany\\
        and\\
        Center for Applied Space Technology and Microgravity\\
        University of Bremen\\
        Am Fallturm, D-28359 Bremen, Germany
}

\date{}

\maketitle

\begin{abstract}
\noindent
Theories in physics usually do not address ``the present''
or ``the now''. However, they usually have a precise notion 
of an ``instant'' (or state). I review how this notion 
appears in relational point mechanics and how it suffices 
to determine durations - a fact that is often ignored in 
modern presentations of analytical dynamics.  An analogous 
discussion is attempted for General Relativity. Finally we 
critically remark on the difference between relationalism 
in point mechanics and field theory and the problematic 
foundational dependencies between fields and spacetime. 

This contribution is based on a talk delivered at the workshop 
\emph{The Forgotten Present. A Quest for a Richer Concept of Time},
held at the Parmenides Foundation at Munich-Pullach from 
April 29th - May 2nd 2010. This written up version will be 
published in the Volume \emph{The Forgotten Present},
edited by Thomas Filk and Albrecht von M\"uller, to be published 
by Springer Verlag 2013. 
\end{abstract}
\bigskip

 \begin{footnotesize}
 \setcounter{tocdepth}{3}
 \tableofcontents
 \end{footnotesize}
\newpage
\section{Introduction}
All known fundamental physical laws are of \emph{dynamical} type. 
Without exception, they are all required to provide answers 
for \emph{initial-value problems}. This means the following: If we 
specify the state of a physical system the laws allow us to 
deduce further states that are usually interpreted as lying to 
the future, or past, or both, of the initially given one.
Except for General Relativity, this is formally achieved by 
labelling the states by an external parameter $t$ that - without 
further justification - is interpreted as ``time'' (whatever 
this means). In this contribution I wish to point out that this
parameter may be eliminated and that measures of duration can
be read off the sequence of states obtained from the dynamical 
laws. 

In the traditional formulation, an initial-value problem is 
said to be \emph{well posed} if and only if the determination 
of the future (and possibly past) states is unique, and 
continuously dependent on the initial state. The last condition 
means that if we sufficiently restrict the variation of the 
initial state we can let the evolution vary less than any given 
bound.  These conditions are not only satisfied in Newtonian 
mechanics, which serves as a paradigmatic example in this respect,  
but also in the mathematically and conceptually and most 
complicated theories, like Einstein's theory of General Relativity.
Albert Einstein, as well as David Hilbert, wrote down the field 
equation of General Relativity in November 1915. But only in 
the late 1950s did mathematicians succeed to prove that it 
indeed allowed for \emph{well posed initial-value problems}. 
Had this turned out to be false this would have possibly led
physicists to abandon General Relativity, despite all its 
other convincing features. To allow for a well posed initial-value
problem is presumably the single most important sanity check 
for any candidate fundamental dynamical law in physics.

This is not restricted to classical laws and classical determinism. 
The fundamental dynamical law in Quantum Mechanics, Schr\"odinger's
equation, also allows for well posed initial-value problems. The 
quantum-mechanical state evolves according to this equation just as 
deterministically and continuously as the state in Newtonian mechanics
does according to Newton's or Hamilton's equations. The typical 
quantum-mechanical indeterminacy, that distinguishes it so 
drastically from classical mechanics does not concern the 
evolution of states, it concerns the relation of states to 
observable features of the system under consideration. But this shall 
not be the issue we address here. Therefore we will restrict 
attention to classical (i.e. non quantum) laws. Our concern 
is the problem of how to characterise, in a physically meaningful 
way, data that suffice to determine the evolution and how to find 
a measure of duration merely from that data.  
\newpage
\section{Newtonian Mechanics}
Newton's famous third law is written in standard modern text-book 
language as 
\begin{equation}
\label{eq:NewtonThirdLaw}       
m\ddot{\vec x}=\vec F\bigl(t,{\vec x}(t),\dot{\vec x}(t)\bigr)\,.
\end{equation}
In this form it is meant to apply to an idealised object called
\emph{mass point}. This may the thought of as extensionless 
object (``point'') of position $\vec x$ and mass value $m$. 
A single overdot denotes the derivative with respect to the 
parameter (``time'') $t$ (i.e. the rate of change of the 
dotted quantity) and a double overdot the second ``time'' derivative.  
Finally, the right-hand side denotes the force, $\vec F$, which 
in the case of just one particle is supposed to be externally 
specified and possibly dependent on $t$, the instantaneous
position $\vec x(t)$ of the particle and its instantaneous velocity 
$\dot{\vec x}(t)$. Given the function $\vec F$, Newton's equation 
has a unique solution once the initial position and initial velocity 
of the particle is specified. The solution is the function 
$t\rightarrow\vec x(t)$ that assigns a unique position $\vec x$ 
in space to each value $t$. That is the standard text-book
presentation, except that $t$ is from the start always referred 
to as time (Newtonian time). 

Equation (\ref{eq:NewtonThirdLaw}) tells us that an initial datum
that suffices to predict the future is the position and velocity
\emph{at the initial reading of time}. The initial reading of time is a 
particular value of the parameter $t$ that \emph{represents} time, 
namely that value that represents the initial moment. This is 
achieved via a \emph{clock}. A clock is a another physical systems 
that also obeys an equation of the form (\ref{eq:NewtonThirdLaw}) 
for the pointer variable $p$ as function of parameter $t$. 
Whereas $t$ is not directly observable, $p$ is. Given $p(t)$ we 
may invert this relation and express $t$ as function of $p$. 
This is possible if $p$ is strictly monotonous in $t$. Systems 
for which this is not the case would not count as clocks. We 
then eliminate $t$ in $\vec x(t)$ in favour of $p$ and obtain a 
function $\vec x(p)$. This function expresses a relation between 
the clock's pointer position $p$  and the particle's position 
$\vec x$. That relation is observable because $p$ \emph{as well as}
$\vec x$ are observable. This is in contrast to $\vec x(t)$ where $t$
is not observable. The elusive ``initial time'' is then that reading 
of $p$ at which we release the particle. This, in essence, is the 
idea of \emph{ephemeris time}~\cite{Clemence:1948}.

But what happens if there is no obvious way to single out a system as 
``clock''. For example, imagine we are given $n+1$ (we say $n+1$ 
rather than $n$ for later notational convenience) mass points moving 
about under the action of their own pairwise gravitational attraction. 
No ``clocks'' or background reference systems against which the 
motions of the particles could be measured are given to us. The only 
thing we can measure are the $\tfrac{1}{2}n(n+1)$ instantaneous 
relative distances between pairs of points. Could we still 
ascertain the validity of Newton's laws of mechanics? This is 
a relevant question since the situation depicted is basically 
just that astronomers have to face. And yet it took almost 200 
years from the writing of Newton's Principia until physicists 
and mathematicians first answered this question (of which 
Newton was fully aware) with sufficient clarity.

The basic question that needs to be answered is how we can 
construct Newton's absolute space and time from observations of
relational quantities alone, for it is only with respect to special 
spatial reference frames and special measures of time that 
Newton's equations are valid. Following Ludwig Lange~(1863-1936) \cite{Lange:1885}, these spatial reference frames are called 
\emph{inertial systems} and the special measures of time 
\emph{inertial timescales}. In this work Lange showed how to 
characterise the inertial system and timescale by means of 
continuous monitoring the motion of three force-free particles.
We shall not discuss Lange's argument here, which has been reviewed
elsewhere~\cite{Giulini:2002b}. Rather, we focus on an alternative 
approach initiated a year earlier by James Thomson~(1822-1892), 
the elder brother of William Thomson~(1824-1907) [better known 
as Lord Kelvin], who in 1884 wrote the following~\cite{Thomson:1884}:
\begin{quote}
``The point of space that was occupied by the centre of the ball at 
any specified past moment is utterly lost to us as soon as that moment 
is past, or as soon as the centre has moved out of that point, having 
left no trace recognisable by us of its past place in the universe of 
space. There is then an essential difficulty as to our forming a 
distinct conception either of rest or of rectilinear motion 
through unmarked space. [...] We have besides no preliminary 
knowledge of any principle of chronometry, and for this additional 
reason we are under an essential preliminary difficulty as to 
attaching any clear meaning to the words \emph{uniform rectilinear 
motion} as commonly employed, the uniformity being that of 
equality of spaces passed over in equal times.''
\end{quote}
This was immediately rephrased into a mathematical problem by 
Peter Guthrie Tait (1831-1901)~\cite{Tait:1884}:
\begin{quote}
``A set of points move, Galilei wise, with reference to a system of 
co-ordinate axes; which may, itself, have any motion whatever. 
From observation of the \underline{\emph{relative}} positions of 
the points, merely, to find such co-ordinate axes.''
\end{quote}
This is precisely the problem we set above in the simpler case 
of \emph{free} point particles. So suppose we are given some 
number of point particles that move about freely, i.e. there 
is no mutual attraction or repulsion due to any force, 
and suppose this motion does obey Newtons law with reference 
to some unknown inertial reference system and inertial 
timescale. How can we reconstruct these by merely observing 
the relative distances of the points? How many points and 
how many snapshots do we need to accomplish that? 

\subsection*{Reconstructing Absolute Space and Time}
Tait's answer to the above question, given in the same paper 
\cite{Tait:1884}, is as follows: We wish to reconstruct the inertial 
system and timescale from an unordered \emph{finite} number of snapshots 
(``instances'') of instantaneous relative spatial configurations. For this 
we consider $n+1$ mass-points $P_i$ ($0\leq i\leq n$) moving 
inertially, i.e. without internal and external forces, in flat 
space. Their trajectories are represented by $n+1$ functions 
$t\mapsto\vec x_i(t)$ with respect to some, yet unspecified, 
spatial reference frame and timescale. The only directly 
measurable quantities at this point are the $n(n+1)/2$ 
instantaneous mutual separations of the particles.
We now proceed in the following nine elementary steps: 

\begin{enumerate}
\item
The instantaneous mutual separations are given by $n(n+1)/2$
positive real numbers per label $t$. This is equivalent to 
giving their squares: 
\begin{equation}
\label{eq:TaitsSol1}
R_{ij}:=\Vert\vec x_i-\vec x_j\Vert^2\qquad
\mathrm{for}\quad  0\leq i<j\leq n\,.
\end{equation}
\item
The knowledge of the $n(n+1)/2$ squared distances, $R_{ij}$, is, 
in turn, equivalent to the $n(n+1)/2$ inner products
\begin{equation}
\label{eq:TaitsSol2}
Q_{ij}:=(\vec x_i-\vec x_0)\cdot(\vec x_j-\vec x_0)\qquad
\mathrm{for}\quad  1\leq i\leq j\leq n\,,
\end{equation}
as one sees by expressing one set in terms of the other by the 
simple linear relations (no summation over repeated indices here): 
\begin{subequations}
\label{eq:TaitsSol3}
\begin{alignat}{4}
\label{eq:TaitsSol3a}
& R_{ij}\,&&=\,Q_{ii}+Q_{jj}-2Q_{ij}\qquad
&&\mathrm{for}\quad && 1\leq i<j\leq n\,,\\
\label{eq:TaitsSol3b}
&R_{i0}\,&&=\,Q_{ii}\qquad
&&\mathrm{for}\quad && 1\leq i\leq n\,,\\
\label{eq:TaitsSol3c}
&Q_{ij}\,&&=\,\tfrac{1}{2}\bigl(R_{i0}+R_{j0}-R_{ij}\bigr)\qquad
&&\mathrm{for}\quad && 1\leq i\leq j\leq n\,.
\end{alignat}
\end{subequations}
\item
We now seek an inertial system and an inertial timescale, with 
respect to which all particles move uniformly on straight 
lines. Correspondingly, we assume   
\begin{equation}
\label{eq:TaitsSol4}
\vec x_i(t)=\vec a_i+\vec v_i t\qquad
\mathrm{for}\quad 0\leq i\leq n
\end{equation}
hold for some \emph{time-independent} vectors $\vec a_i$ and 
$\vec v_i$. 
\item
The 11-parameter redundancy by which such inertial systems and 
timescales are defined is given by 
\begin{itemize}
\item[a)]
spatial translations: $\vec x\mapsto\vec x+\vec a$, 
$\vec a\in\reals^3$, accounting for three parameters,
\item[b)]
spatial boosts: $\vec x\mapsto\vec x+\vec vt$, 
$\vec v\in\reals^3$, accounting for three parameters,
\item[c)]
spatial rotations: $\vec x\mapsto\matr{R}\cdot\vec x$, 
$\matr{R}\in\lieg{O}(3)$ (group of spatial rotations, including 
reflections), accounting for three parameters, 
\item[d)]
time translations: $t\mapsto t+b$, 
$b\in\reals$, accounting for one parameter, and
\item[e)]
time dilations: $t\mapsto at$,
$a\in\reals-\{0\}$, accounting for one parameter.  
\end{itemize}
The redundancies a) and b) are now eliminated by assuming 
$P_0$ to rest at the origin of our spatial reference frame. 
We then have, assuming \eqref{eq:TaitsSol4},
\begin{equation}    
\label{eq:TaitsSol5}
Q_{ij}(t)=\vec x_i(t)\cdot\vec x_j(t)=
\vec a_i\cdot\vec a_j+
t\,(\vec a_i\cdot\vec v_j+\vec a_j\cdot\vec v_i)
+t^2\,\vec v_i\cdot\vec v_j\,.
\end{equation}
\item
Measuring the mutual distances, i.e. the $Q_{ij}$, at $k$ different 
values $t_a$ ($1\leq a\leq k$) of $t$ we obtain the $kn(n+1)/2$ 
numbers $Q_{ij}(t_q)$. From these we wish to determine the following 
unknowns, which we order in four groups: 
\begin{itemize}
\item[1)]
the $k$ times $t_a$,
\item[2)]
the $n(n+1)/2$ products $\vec a_i\cdot\vec a_j$,
\item[3)]
the $n(n+1)/2$ products $\vec v_i\cdot\vec v_j$, and
\item[4)]
the $n(n+1)/2$ symmetric products 
$\vec a_i\cdot\vec v_j+\vec a_j\cdot\vec v_i$.
\end{itemize}
\item
The arbitrariness in choosing the origin and scale of the 
time parameter $t$, which correspond to the points d) 
and e) above, can, e.g., be eliminated by choosing $t_1=0$ and 
$t_2=1$. Hence the first group has left the $k-2$ unknowns $t_3,\dots,t_k$. 
The last remaining redundancy, corresponding to the spatial 
rotations in point c), is \emph{almost} eliminated by choosing 
$P_1$ on the $z$ axis and $P_2$ in the $xz$ plane. This suffices   
as long as $P_0,P_1,P_2$ are not collinear. Otherwise we choose 
three other mass points for which this is true. Here we exclude 
the exceptional case where all mass points are co-linear. 
We said that this `almost' eliminates the remaining redundancy, 
since a spatial reflection at the origin is still possible.
\item
Tait's strategy is now as follows: for each instant in time $t_a$ 
consider the $n(n+1)/2$ equations (\ref{eq:TaitsSol5}). 
There are $k-2$ unknowns from the first and $n(n+1)/2$
unknowns each from groups 2), 3), and 4). This gives a total of 
$kn(n+1)/2$ equations for the $k-2+3n(n+1)/2$ unknowns. 
The number of equations minus the number of unknowns is 
\begin{equation}
\label{eq:TaitsSol6}
(k-3)n(n+1)+2-k\,.
\end{equation}
This is positive if and only if $n\geq 2$ and $k\geq 4$.
Hence the minimal procedure is to take four snapshots ($k=4$) of three 
particles ($n=2$), which results in 12 equations for 11 unknowns.
\item
Recall that we assumed the validity of Newtonian dynamics and 
that the given trajectories correspond to force-free particles. 
This implies the existence of inertial systems and hence also 
the existence of solutions to the equations above. For positive 
(\ref{eq:TaitsSol6}) the equations determine the $3n(n+1)/2$ 
unknowns in groups 2) - 4) which, in turn, determine the $6n-3$ 
free components of $\vec a_i$ and $\vec v_i$ up to an overall 
sign, since $3n(n+1)/2\geq 6n-3$ if and only if $n\geq 2$. 
Note that we have $6n-3$ rather than $6n$ free components for 
$\vec a_i$ and $\vec v_i$, since we already agreed to put $P_1$ 
on the $z$ axis, which fixes two components of 
$\vec a_1$ and $\vec v_1$ each, and $P_2$ in the $xz$ plane, which 
fixes one component of $\vec a_2$ and $\vec v_2$ each. Note also 
that we cannot do better than determining the $\vec a_i$ and 
$\vec v_i$ up to sign, since the $Q_{ij}$ are homogeneous functions 
of \emph{second} degree in these variables. 
\item
Once the $2n$ vectors $\vec a_i$ and $\vec v_i$ are obtained,
so is clearly the inertial system (up to orientation) and the 
inertial timescale. This is as far as Tait's solution to 
Thomson's problem goes.
\end{enumerate}

One remarkable thing about Tait's solution is that the spatial
inertial system and the inertial timescale are determined 
together. However, this is really not surprising: 
The mathematical problem of calculating the $k$ labels $t_a$ representing 
``instants'' cannot be separated from the characterisation 
of the instants themselves. In this sense it might be said 
-- following Julian Barbour \cite{Barbour:1994a} -- that 
instants are not to be located in time, but that time is 
rather to be found in instants. Thus it seems that the 
philosophical discussion concerning the reality of time 
(see e.g. \cite{SEoP-Time} for an up-to-date account)  
is then really a discussion concerning the reality of 
instants. But in point mechanics instants are  relational 
configurations the reality of which cannot be doubted
without mocking the theory. 

\subsection*{Mechanics without parameter-time}
If time can be read off instances, as claimed above, we should,
at least in principle, be able to altogether eliminate the parameter 
$t$ from the laws. How does the $t$-less version of Newtonian 
mechanics look like? One answer has been well known for a long time,
albeit in a somewhat hybrid form in which the absolute positions 
in space still feature. It goes under the name of Jacobi's principle, 
after Carl Gustav Jacobi (1804-1851). It takes the form of a geodesic principle in configuration space. That means, it determines the 
physically realised paths in configuration space between any pair 
$(\q_i,\q_f)$ of given points to be that of shortest length. 
Here ``length'' is measured in some appropriate metric that 
encodes the essential dynamical information. 

Note that the parameter $t$ plays no r\^ole: its value at the 
initial and final point need not be specified. Rather, the 
measure of inertial time elapsed between the initial and final 
configuration can be calculated \emph{after} the dynamical 
trajectory has been determined through the geodesic principle. 
Let there be $n$ mass points whose positions are 
$(\vec q_1,\cdots,\vec q_n)=:\q$, moving under the influence 
of a potential $V(\q)$. The configuration space is  $\reals^{3n}$ 
and its Riemannian metric, with respect to which the physically 
realised trajectories of constant Energy $E$ are geodesics, is 
given by $g=(E-V)T$, where $T$ in the positive-definite 
bi-linear form that appears in the expression for the kinetic 
energy (``kinetic-energy metric''). The inertial time that has 
elapsed along the length-minimising trajectory between $\q_i$ 
and $\q_f$ is then given by   
\begin{equation}
\label{eq:JacobiTimeSpan-1}
\Delta t(\q_i,\q_f)=
\int_{\q_i}^{\q_f}\sqrt{\frac{T\bigl(d\q/d\lambda,d\q/d\lambda\bigr)}{E-V(\q)}}\,d\lambda\,.
\end{equation}
This may be understood as saying that time has to be chosen in such
a fashion so as to lead to the standard form of energy conservation.
Indeed, from \eqref{eq:JacobiTimeSpan-1} we get
\begin{equation}
\label{eq:JacobiEnergyLaw}
E=T\bigl(d\q/dt\,,\,d\q/dt\bigr)+V(\q)\,.
\end{equation}
Note that \eqref{eq:JacobiTimeSpan-1} only depends on the pair 
$(\q_i,\q_f)$ and not on the way we parametrise the path. 
Hence the choice of the parameter $\lambda$ is arbitrary. 
Therefore we have a well defined map 
\begin{equation}
\label{eq:JacobiTimeSpan-2}
\Delta t: \reals^{3n}\times\reals^{3n}\rightarrow\reals_+
\end{equation}
which, for given energy $E$, assigns to each pair of points 
in the configuration space the inertial-time duration of the 
physical journey connecting them. As we will discuss next, there 
is a certain analog to Jacobi's Principle in General Relativity,
with some additional issues arising due to the fact that 
the fundamental mathematical entities being fields rather than 
point-particles. Finally we point out that there is a 
generalisation to Jacobi's principle in models of point 
mechanics without absolute space. In these models only the 
instantaneous relative distances enter the laws and the 
time lapse can again be calculated from the dynamical 
trajectories. First attempts were Reissner's 
(1874-1967)~\cite{Reissner:1914} and 
Schr\"odinger's~\cite{Schroedinger:1925}, 
with the full ``relativisation'' of time being achieved 
only much later in \cite{Barbour.Bertotti:1982}.
See also \cite{Barbour.Pfister:MachsPrinciple} for more on the 
modern context and translations of the papers by Reissner, 
Schr\"odinger etc.

\section{General Relativity}
Einstein's equations are equations for entire spacetimes, 
that is, pairs $(M,g)$ where $M$ is a four-dimensional 
differentiable manifold endowed with a certain geometric 
structure called \emph{Lorentzian metric}, which is here 
represented by $g$. Given such a pair $(M,g)$ and a 
specification of certain aspects of physical matter, 
it makes unambiguous sense to say that $(M,g)$ does, 
or does not, satisfy Einstein's equations. No external 
notion of time enters the picture at this stage. This,
clearly, is for good reasons: Spacetimes do not 
evolve (in ``time'' external to them); they simply 
are! In addition, no conditions concerning structures 
internal to $(M,g)$ need to be imposed, such as 
sequential ordering of substructures (to be interpreted 
as ``instants''), absence of closed timelike curves (i.e. 
journeys into ones own past), or causal evolution of 
geometry. On the other hand, Einstein's equations are 
\emph{compatible} with the \emph{additional imposition} 
of such structures. It required the hard work of 
mathematicians of many years to show that a reasonable 
set of such additional conditions exist which ensure 
that Einstein's equations allow for well posed initial 
value problems in the sense explained above. 
\begin{figure}[htb]
\centering\includegraphics[width=0.7\linewidth]{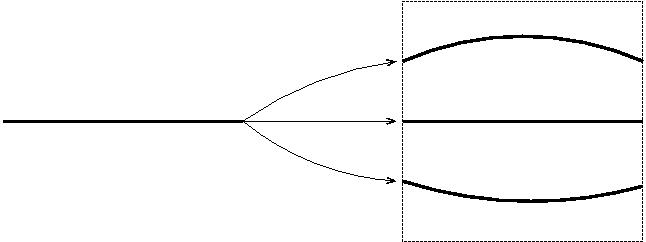}
\put(-200,50){\small $\Sigma$}
\put(-125,50){\small $\mathcal{E}_t$}
\put(-125,30){\small $\mathcal{E}_{t'}$}
\put(-125,70){\small $\mathcal{E}_{t''}$}
\put(-50,50){\small $\Sigma_t$}
\put(-50,20){\small $\Sigma_{t'}$}
\put(-50,84){\small $\Sigma_{t''}$}
\put(-90,80){\small $M$}
\caption{\label{fig:EmbeddingOfSpaces}\small
Spacetime, $M$, is foliated by a one-parameter family of 
embeddings $\mathcal{E}_t$ of the 3-manifold $\Sigma$ 
into $M$. Here $t$ is a formal label without direct 
physical significance. $\Sigma_t$ is the image in $M$ 
of $\Sigma$ under $\mathcal{E}_t$. Each such $\Sigma_t$ 
is an \textbf{instant}.}  
\end{figure}
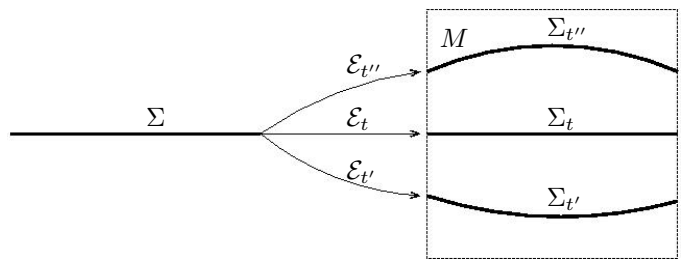

\begin{figure}
\begin{minipage}[b]{0.48\linewidth}
\centering\includegraphics[width=.68\linewidth]{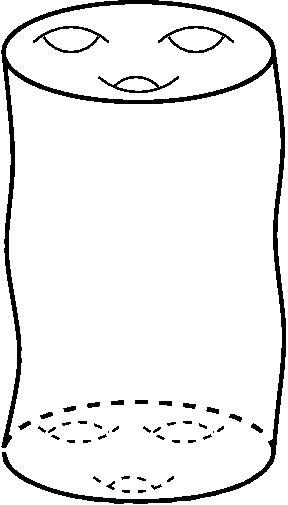}
\end{minipage}
\hfill
\begin{minipage}[b]{0.49\linewidth}
\centering\includegraphics[width=0.9\linewidth]{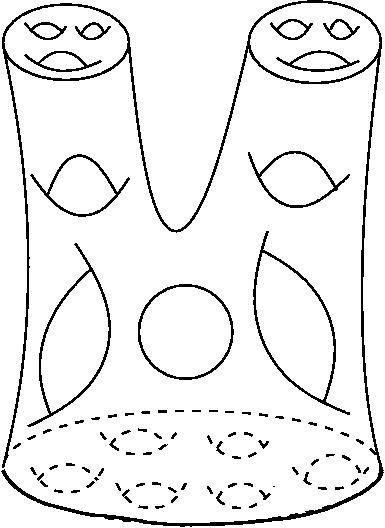}
\end{minipage}
\put(-295,18){\footnotesize initial space}
\put(-295,180){\footnotesize final space}
\put(-292,85){\footnotesize spacetime}
\put(-110,21){\footnotesize initial space}
\put(-107,210){\footnotesize two final}
\put(-112,203){\footnotesize components}
\put(-107,196){\footnotesize of space}
\caption{\label{fig:TwoDifferentSpacetimes}\small
Schematic rendering of spacetimes. The one 
on the left may be viewed as time evolution of space.
Time runs upwards and space corresponds to the horizontal 
sections, here depicted by a 3-holed surface. 
In the spacetime on the right an initial connected space
at the bottom, represented by a single 6-holed surface, 
evolves into two 3-holed pieces. This spacetime cannot be 
viewed as time evolution of a single space and shall 
be excluded from the discussion.}
\end{figure}
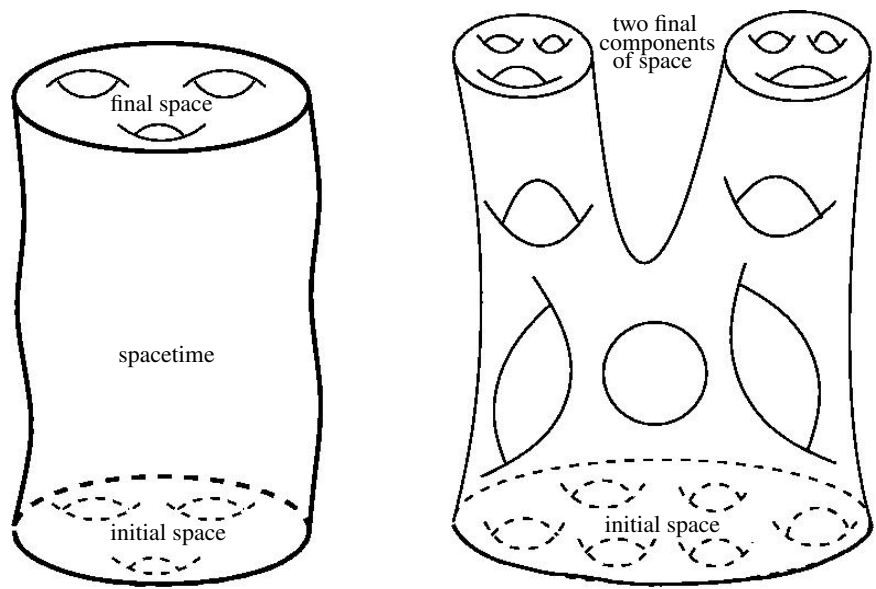

In particular, these conditions ensure that the 
spacetime can be thought of as the history of space.
In a loose mathematical sense this means that 
spacetime is a staking of spaces, each one being an 
instant. More precisely, spacetime is foliated by 
a one-parameter family of embeddings of space into
spacetime. This is schematically represented in 
Figure\,\ref{fig:EmbeddingOfSpaces}. 
For that to make mathematical sense we must be sure 
that a single space, $\Sigma$,  suffices to foliate 
spacetime. Its geometry may change from leaf to leaf, 
but not its essential properties as differentiable 
manifold, for otherwise we could not speak of \emph{its} 
evolution. In particular this means that its topological 
properties are preserved during evolution, like its 
connectedness and its higher topological invariants; see
Figure~\ref{fig:TwoDifferentSpacetimes}.   

\begin{figure}[htb]
\centering\includegraphics[width=0.54\linewidth]{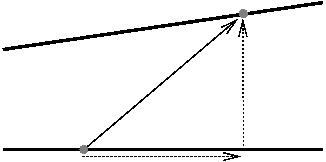}
\put(-180,13){\small $\Sigma_{t}$}
\put(-180,78){\small $\Sigma_{t+dt}$}
\put(-145,15){\small $p$}
\put(-51,94){\small $p'$}
\put(-100,-5){\small $\beta$}
\put(-45,45){\small $\alpha n$}
\put(-110,50){\small $\frac{\partial}{\partial t}$}
\caption{\label{fig:LapseShift}\small
There is a large ambiguity in moving from an initial 
space-slice $\Sigma_t$ ``forward in time''. 
For $q\in\Sigma$ the image points $p=\mathcal{E}_t(q)$ 
and $p'=\mathcal{E}_{t+dt}(q)$ are connected by the 
vector $\partial/\partial t\vert_p$ whose components 
tangential and normal to $\Sigma_t$ are $\beta$ 
(three functions) and $\alpha n$ (one function) 
respectively. Hence there is a four-function worth 
of ambiguity to move $\Sigma_t$ in a given ambient 
spacetime.}   
\end{figure}

One of the fundamental difficulties with the notion 
of spacetime as history of space is its inherent 
redundancy: There are many ways to describe one and 
the same spacetime as the evolution of space. This 
is explained in Figure~\ref{fig:LapseShift}. This 
means that if we cast Einstein's equations into the 
form of evolution equations for ``space'', we cannot
expect unique solutions,  contrary to what is usually 
required for well posed initial-value problems. The 
point here is that the non-uniqueness is not arbitrary.
It is precisely of the amount that accounts for the 
different ways to move space through a \emph{fixed} 
spacetime, no more and no less. This is closely related to 
the infamous ``Hole Argument''~\cite{SEoP-HoleArgument}.
That relaxation of the uniqueness requirement is familiar 
from so-called ``gauge-theories'' and does not imply any 
renunciation from determinism of fundamental laws, at 
least as long as the degree of arbitrariness in the 
analytical expression of the evolution is under complete 
mathematical control. Physical configurations are then
taken to be the equivalence classes under the relation 
that identifies any two apparently different evolutions 
that give rise to the same spacetimes (more precisely: 
diffeomorphism-class of spacetimes). 

\subsection*{The Chronos principle}
Modulo the difficulties just mentioned, we may ask 
whether we can extract a notion of time merely from 
the information of instants. An instant here is a 
spatial configuration, that is a pair $(\Sigma,h)$,
where $\Sigma$ is a 3-dimensional manifold and $h$ 
is a Riemannian (i.e. positive definite) metric. 
One obvious question concerning Einstein's equations 
is this: given two instants $(\Sigma_1,h_1)$ and 
$(\Sigma_2,h_2)$, can we associate a measure of 
time by which they are apart if we assume that 
both 3-geometries occur in a spacetime that satisfies 
Einstein's equations. This is known as the 
``sandwich conjecture'' in General Relativity and 
known to fail in many known examples which are, however, 
of special symmetry that renders the problem singular. 
For example, it is obvious that specifying any two flat 
3-slices in Minkowski space does not give us any information
on their separation. Similarly, it has been shown that 
in the spherically symmetric case a similar 
underdetermination prevails~\cite{Murchadha.Roszkowski:2006}.
On the other hand, it has been an old hope that a 
suitable analog of Jacobi's principle, and in particular
formula \eqref{eq:JacobiTimeSpan-1}, is also valid in 
General relativity. This has been first proposed in the 
classic and well known paper~\cite{Baierlein.Sharp.Wheeler:1962}
of 1962. An apparently less well known contribution 
appeared 12 years later, in which the following 
``Chronos Principle'' in General Relativity was 
proposed, according to which time is a measure for 
the distance of instantaneous configurations 
(instants)~\cite{Christodoulou:1975}. Moreover, 
it was asked in \cite{Christodoulou:1975} whether 
such measures existed such that one would not have 
to know the entire spatial configuration in order 
to determine the time span. 

\begin{quote}
``This postulate contains the statement that it 
is not necessary to look at the change in 
configuration of the entire universe to measure 
time. It is sufficient to measure the change 
in configuration of only a localized region 
of the universe, and one is assured that the 
local time thus obtained will be equal to that 
of any other region, and indeed equal to the 
global time.'' (\cite{Christodoulou:1975}, p.\,76)         
\end{quote}
It is this localisation property that renders 
this reading of time from instants physically 
viable. Let us therefore see how it can be satisfied. 
The answer, quite surprisingly, leads more or 
less directly to General Relativity. We shall 
give the argument in a slightly  simplified form.

As already stated, Einstein's equations can be cast into 
evolutionary form. In that form one may identify a 
kinetic-energy metric, just like in point mechanics. 
It reads:   
\begin{equation}
\label{eq:WDW-Distance}
ds^2=\int_\Sigma d^3x\ G^{ab\,nm}[h(x)] dh_{ab}(x)dh_{nm}(x)
\end{equation} 
where $G^{ab\,nm}[h(x)]$ is a certain expression that 
depends on the metric tensor $h$ of space but not on its 
derivatives (ultralocal dependence). It is sometimes 
called the Wheeler-DeWitt metric. The measure of time 
will be obtained by a rescaling of the kinetic-energy 
metric, just like in \eqref{eq:JacobiTimeSpan-1}. 
Hence one writes 
\begin{equation}
\label{eq:ChronosTime}
d\tau^2=\frac{ds^2}{\int_\Sigma d^3x\ R(x)}\,.
\end{equation}
Here $R$ must be a scalar function of the spatial 
metric $h$. The simplest non-constant such function is 
the scalar curvature, which depends on $h$ and its 
derivatives up to order 2. The condition that the measure 
of time be compatible with arbitrarily fine localisation 
$\Sigma\rightarrow U\subset\Sigma$ leads requires the 
integrands in the numerator and denominator of 
(\ref{eq:ChronosTime}) to be proportional. Without 
loss of generality we can take this constant of 
proportionality (which cannot be zero) to be $1$
(this just fixes the overall scale of physical time) 
and obtain 
\begin{equation}
\label{eq:ChronosTimeLocProp}
G^{ab\,nm}[h(x)] \frac{dh_{ab}(x)}{d\tau}\frac{dh_{nm}(x)}{d\tau}-R[h](x)=0\,.
\end{equation}
This is a well known formula (the so-called Hamiltonian 
constraint) in General Relativity. Hence Relativity just 
satisfies the localisation property with the simplest 
conceivable local rescaling function $R$. Finally, 
physical time is now given in terms of 3-dimensional 
geometric quantities by a Jacobi-like formula, which 
is just the analog of \eqref{eq:JacobiTimeSpan-1} 
in the case $E=0$:
\begin{equation}
\label{eq:TimeFormula}
\Delta\tau(\h_i,\h_f)=\int_{\h_i}^{\h_f}
\sqrt{\frac{
G\bigl(d\h/d\lambda,d\h/d\lambda\bigr)}
{-R\bigl[\h(\lambda)\bigr]}}\ d\lambda
\end{equation}

\section{Conclusions and open issues}
Following \cite{Barbour:1994a} we tried to argue that the 
notion of ``time from instances'' in inherent in classical
point mechanics as well as General Relativity. We also 
saw that in General Relativity that notion of time is 
not as hopelessly global as one  might have feared. In fact,
one can argue that General Relativity just realises the 
simplest \emph{localisable} notion of that sort of time. 

But there are also points that remain open (to me): 
\begin{enumerate}
\item
Solutions to dynamical equations of motion in the form 
of (generalised) geodesic principles are subsets of 
(dynamically realised) configurations in the space of 
(kinematically possible) ones. These subsets are delivered 
to us in the form of unparametrised curves. So, even 
though the parameter does not matter, the structure of 
a one-dimensional sub-continuum remains. In particular 
one (or two) preferred orderings are selected. What is 
the significance of that? What makes us experience this 
solution configurations according to this order? 
\item
Can we, on the space of 3-geometries, characterise a 
function that structures it according to some definition 
of geometric entropy? How would its gradient flow be 
related to the dynamics of General Relativity?
\item
Suppose the spacetime we live in did not allow for any 
symmetries and were sufficiently generic, so as to not 
allow for two \emph{different} isometric embeddings of 
any of its possible 3-geometries.  (Such spacetimes exist
and are, intuitively speaking, the generic case, though 
their degree of generality or naturalness is not easy to 
characterise mathematically.) This means that each 
instant would have its unique place in spacetime. Would 
this count as a perfect representation of the ``Now'' 
in a physical theory (here General Relativity), or 
could/should we ask for more? 
\end{enumerate}

Finally I wish to comment on the transition of point mechanics 
to field theory. In point mechanics, the requirement to only 
employ purely relational quantities is met by eliminating all 
explicit reference to absolute space and time. This has been 
gradually achieved in the papers of Reissner, Schr\"odinger, 
and Barbour \& Bertotti. But what is the precise analog of that 
requirement in field theory? A standard answer to this is that 
the theory should be \emph{background independent}. The intended 
meaning of that phrase is that the theory should not employ 
structures which are not dynamically active. Closer inspection 
shows that it is quite hard to translate this intended meaning 
into a clear mathematical condition~\cite{Giulini:2007b}. The problem 
is that whatever the mathematical formulation is, it seems quite 
easy to turn it into an equivalent one by some formal rewriting 
that renders it (formally) background independent. It is often taken 
for granted that the requirement of \emph{diffeomorphism invariance} 
(also known as ``general covariance'') is sufficient, because that 
would deprive spacetime points of their independent individuality.
This is true to some extend, but it seems not to go as far as one 
might have hoped for. Modern (quantum-)field theory does not 
get rid of space and time.  

Markus Fierz was deeply concerned about the problematic 
relation between spacetime and fields. In a remarkable 
letter of October 9-10th 1951 to Wolfgang Pauli\footnote{There exist two 
versions of this letter, one from October 9th and one from 
October 10th. Here we quote from the fist only.} he wrote 
(\cite{Pauli:SC}, Vol.\,IV, Part\,I, Doc.\,1287, p.\,379)
\begin{quote}
``There exist [in classical physics--DG] solutions [to field equations--DG]
with empty domains, that is, emptied from all fields. Hence one needs a 
theory of space which is independent of what fills space. There is the geometry of space and the laws of things in space. [...] 

Space is still absolute in Relativity Theory insofar as one may 
characterise it without referring to its `content', and because it 
may even exist without any content. [...] 

In a [hypothetical--DG] full Theory of Quantum Fields, in which the act
of observation and the possibility to localise are described correctly, 
it should not be necessary to introduce space separately.
Opposite to what Einstein hoped, the laws of space should follow
from the laws of Nature (not the laws of Nature from geometry).
But this can only be hoped for if there is no such thing as empty
space, that is, if you cannot clear [ausr\"aumen] space. Fields are 
not in space, they span space. Space is not a geometric idea [Gedankending], 
it is a certain aspect of the world.''

In this sense, space in Relativity Theory is absolute and this is
why Einstein suggested to call it aether. In a proper field theory
the theory of localisation should deliver a theory of space. Space
should somehow be `created' by test bodies and hence be a function 
of the observer in a much deeper sense than in Relativity
Theory.''
\end{quote}
Pauli replied on October\,13 in a way that would also be typical 
for many modern relativists (\cite{Pauli:SC}, Vol.\,IV, Part\,I, Doc.\,1289, 
pp.\,385-386):
\begin{quote}
``Your wording does not do justice to Relativity Theory, which is
just an attempt to connect geometry and laws of nature concerning 
things [Dinge] in the spacetime world. [...] All people happily
proclaim just the opposite to what you said in your letter: namely
‘from now on only the connection of spacetime and things is 
absolute. [...] 

I am quite indignant about this part of your letter, since it 
shows to me that the, compared to me, slightly younger generation 
of physicists (not to speak of the still younger ones!) have 
completely repressed [verdr\"angen] General Relativity - 
and because I know how important Einstein considered this 
point to be. [...]

After this urgent correction (diagnosis: `repression' [Verdr\"angung]!) 
one can ask whether the dependence of space (i.e. spacetime) from the 
things [Dingen] according to General Relativity is sufficient. To pose 
the question already means to negate it. [...] 

I agree that the impossibility to accommodate Einstein's postulate 
(i.e. Mach's original point of view) within General Relativity is a 
deep and significant sign for the inadequacy of classical field physics.''
\end{quote}  
So we see that after his usual grumble Pauli finally agrees at 
least on the existence of a fundamental difficulty, which was, 
after all, well addressed by Fierz' original complaint. Even 
today all candidate theories of quantum gravity make use of 
non-dynamical structures that represent some sort of 
space or spacetime (of various dimensions). Hence I believe Fierz' 
complaint is as as relevant today as it was 60 years ago.  
  
Everyone knows the opening words of Hermann Minkowski's 
(1864-1909) famous address ``Raum und Zeit'', delivered 
in Cologne on September 21st 1908~\cite{Minkowski:1909}:
\begin{quote}
``Gentlemen! The views of space and time which I wish to lay before you have 
sprung from the soil of experimental physics. Therein lies their strength. 
They are radical. Henceforth space by itself, and time by itself, are doomed 
to fade away into mere shadows, and only a kind of union of the two will 
preserve an independent reality.''
\end{quote}
But it seems not to be so well known that Minkowski felt the enormous 
abstraction and possible physical over-idealisation of the concept of 
spacetime \emph{as such}, as he clearly indicated in his 
introduction, before going into the description of what we now call 
``Minkowski space'' (space meaning spacetime).  He wanted his readers 
to understand the points of spacetime as individuated entities:
\begin{quote}
``In order to not leave a yawning void [g\"ahnende Leere], 
we wish to imagine that at every place and at every time 
something perceivable exists. In order to avoid saying `matter'
or `electricity' for that something, I will use the word 
`substance' for it. We focus attention on the substantive 
[substantiellen] world point at $x,y,z,t$ and imagine to be 
capable to recognise this substantive point at any other time''.
\end{quote}
That substantivalist's view of Minkowski spacetime is still inherent 
in its mathematical representation in modern field theory. One sign of 
this is the interpretation of its automorphism group (the Poincar\'e
group) as proper physical symmetries rather than gauge transformations. 
Recall that a proper physical symmetry transforms solutions to dynamical 
equations into solutions, but the transformed solution is considered 
physically different (distinguishable) from the original one. In contrast, 
gauge transformations just connect redundant descriptions of the same 
physical situation.

Individuating spacetime points is natural if we think of 
spacetime to be a geometrically structured \emph{set}. A set, 
by Cantor's definition, consists first of all of a set 
which may then carry certain geometric structures of 
various complexities.  But recall what according to Cantor's 
definition it already takes to be a set~\cite{CantorMengenlehre1:1895}:
\begin{quote}
``By a \emph{set} we understand any gathering together $M$ of 
determined well-distinguished objects $m$ of our intuition or 
of our thinking (which are called elements of $M$) into a whole.''
\end{quote}
Minkowski's ``substance'' may serve to distinguish events. But is 
that substance not eventually just another physical system obeying 
its own dynamical laws? If so, what kind of ``dynamical law'' can 
that be if there is no non-dynamical substance left with respect 
to which we can define change. Surprisingly -- or perhaps not -- 
this is just the same difficulty that stood at the very beginning 
of modern theories of dynamics. In ``de gravitatione'', written 
well before the Principia, presumably between 1664 and 1673 (the dating is still controversial), Newton said~\cite{Newton:UeberDieGravitation}:

\begin{quote}
``It is accordingly necessary that the determination of
places and thus of local motions is represented in some 
unmoved being of which sort space or extension alone is 
that which is seen as distinct from bodies. [...]

About extension, then, it is probably expected that it is being 
defined either as substance or accidents or nothing at all. 
But by no means nothing, surely, therefore it has some mode of
existence proper to itself, by of which it fits neither to 
substance nor to accident.''
\end{quote}

\vspace{1.5cm}
 
\begin{center}
{\large\bf ``Das noch \"Altere ist immer das Neue''}\\
Wolfgang Pauli
\end{center}

\vspace{1.5cm}

\subsection*{Acknowledgements}
I sincerely thank Albrecht von M\"uller and Thomas Filk for 
several invitations to workshops of the Parmenides Foundation, 
during which I was given the opportunity to present and discuss 
the material contained in this contribution.

\newpage
\bibliographystyle{plain}
\bibliography{RELATIVITY,HIST-PHIL-SCI,MATH,QM,COSMOLOGY}
\end{document}